\documentclass[runningheads]{llncs}
\usepackage[T1]{fontenc}
\usepackage{graphicx}
\usepackage{amsmath}
\usepackage{subcaption}
\usepackage{rotating}
\usepackage{xurl}
\newcommand{\etal}{\textit{et al}.}
\newcommand{\ie}{\textit{i}.\textit{e}., }

\begin{document}
\title{Dismantling Hate: Understanding Hate Speech Trends Against NBA Athletes}

\author{Edinam Kofi Klutse \and
Samuel Nuamah-Amoabeng \and
Hanjia Lyu \and
Jiebo Luo}
\authorrunning{E. Klutse et al.}
\institute{University of Rochester, Rochester NY 14627, USA}
% \institute{University of Rochester, Rochester NY 14627, USA \and
\maketitle              % typeset the header of the contribution
\begin{abstract}
Social media has emerged as a popular platform for sports fans to express their opinions regarding athletes' performance. Fans consistently hold high expectations for athletes, anticipating exceptional performances week after week. This ongoing phenomenon sometimes gives rise to highly negative sentiments, with the worst-case scenario involving the occurrence of hate speech. The National Basketball Association (NBA) is widely recognized as one of the most popular sports leagues globally. However, an unfortunate aspect that has emerged in recent years is the presence of abusive fans within the league. Consequently, the focus of this research is to identify which NBA athletes experience abuse on Twitter and delve deeper into the underlying reasons behind such mistreatment. To address the research questions at hand, the study employs a curated set of keywords to query the Twitter API, gathering a comprehensive collection of tweets that potentially contain hate speech directed toward NBA players. A deep learning classification model is implemented, effectively identifying tweets that genuinely exhibit hate speech. We further use keyword search methods to detect the specific groups that are targeted by hate speech the most and identify topics of hate speech tweets. The findings of our research indicate that certain groups of athletes are particularly vulnerable to hate speech from fans. Notably, high-performing athletes, {\tt Black} athletes, overweight athletes, short athletes, and athletes associated with the {\tt LGBTQ} community are found to be highly susceptible to abusive remarks. Racism, physique shaming, play style, and anti-LGBTQ remarks are the major themes. These findings contribute to a broader understanding of the challenges faced by NBA athletes in the digital space and provide a foundation for developing strategies to combat hate speech and foster a more inclusive environment for all individuals involved in the NBA community.

\end{abstract}
\section{Introduction}

In recent years, professional athletes in the National Basketball Association (NBA) have increasingly expressed their concerns about being subjected to hatred and abuse from fans and media personnel on various social media platforms~\cite{reynolds2019nba}. Among these platforms, Twitter has emerged as a prominent arena where fans can directly engage with players, making it a hotspot for hate speech directed toward NBA athletes. Unfortunately, the prevalence of derogatory language and abusive behavior on Twitter persists despite efforts to combat it~\cite{reynolds2019nba}. Consequently, basketball players in the NBA have become a vulnerable target group for hate speech abuse.

Within this context, it is essential to address two key questions: Who are the athletes experiencing abuse? And what are the underlying reasons behind this mistreatment? In this study, we curate a set of hate speech-related keywords to collect tweets that potentially contain hateful content against NBA players. We then employ a deep learning model to detect hate speech tweets. The keyword search methods are used to detect the specific groups of athletes that are targeted by hate speech. By analyzing the collected data, the study aims to uncover the major themes prevalent in these hate speech tweets. Next, we conduct correlation analysis on a series of players' performance statistics, their demographics, as well as their physical characteristics. Our study seeks to obtain insights into the underlying motivations behind hate speech abuse in the NBA.

\vspace{-2mm}

\section{Method}
\vspace{-2mm}
\subsection{Hate Speech Detection}
Detecting hateful content on Twitter is not a trivial task because users may use certain codes to avoid detection by automated systems~\cite{magu2017detecting,magu2018determining}. Other challenges may include linguistic subtleties, varying definitions of hate speech, and limited access to data for training and testing such systems~\cite{macavaney2019hate}. In our study, we first use keywords to collect tweets that may contain hateful content and then employ a transformer-based language model to perform the final classification.

\subsubsection{Data Collection}
We use Twitter's API - Tweepy, to gather tweets containing potential hate speech targeting NBA players. To collect such tweets, we first compile a list of hate speech-related keywords. Previous research has indicated that online hate speech can stem from various motivations, including but not limited to racial discrimination, gender-based targeting, and body shaming.\footnote{\url{https://www.news24.com/sport/tennis/commentator-dokic-hits-out-at-fat-shaming-trolls-at-australian-open-20230123}} For instance, Powell~\etal~\cite{powell2020digital} found that transgender individuals experience higher rates of digital harassment and abuse overall, and higher rates of sexual, sexuality, and gender-based harassment and abuse, as compared with heterosexual cisgender individuals. By employing this methodology, we aim to gather a dataset that captures the diverse manifestations of hate speech directed at NBA players on social media. In particular, the keyword list is composed of \textit{nigger}, \textit{nigga}, \textit{bitch}, \textit{b*tch}, \textit{n*gg*r}, \textit{fuck}, \textit{bum}, \textit{motherfucker}, \textit{bollock}, \textit{wanker}, \textit{dirty}, \textit{lame}, \textit{bozo}, \textit{faggot}, \textit{pussy}, \textit{f*ck}, \textit{piece of shit}, \textit{sh*t}, \textit{bastard}, \textit{cock}, \textit{gay}, \textit{lesbian}, \textit{fucker}, \textit{fool}, \textit{cunt}, \textit{asshole}, \textit{hate}, \textit{stupid}, \textit{useless}, \textit{fraud}, \textit{cost me}, \textit{owe me}, \textit{lost money}, \textit{liar}, \textit{trash}, \textit{ass}, \textit{overrated}, \textit{flop}, \textit{flopper}, \textit{flopping}, \textit{coward}, \textit{choker}, \textit{choke artist}, \textit{loser}, \textit{choking}, \textit{selfish}, \textit{stat padder}, \textit{ball hog}, \textit{stat pad}, \textit{soft}, \textit{weak}, \textit{retard}, \textit{prick}, \textit{dick}, \textit{dickhead}. The combinations of the names of current NBA players ($n=461$, obtained from \url{basketballreference.com}) and hate speech-related keywords are used to query tweets through Tweepy. In the end, we identify a total of 503,424  tweets of potential hate speech targeting current NBA players.

\subsubsection{Modeling}
A tweet that contains hate speech-related lexicons might be an instance of \textit{offensive language} instead of hate speech which is defined as ``language that is used to express hatred toward a targeted group or is intended to be derogatory, to humiliate, or to insult the members of the group''~\cite{davidson2017automated}. Therefore, we further leverage a transformer-based language model to detect hate speech from the collected tweets. Transformer-based models have demonstrated exceptional performance in text classification tasks across various domains~\cite{chen2021fine,zhang2021monitoring,lyu2022social}. In particular, we first use an open-source hate speech dataset built by Davidson~\etal~\cite{davidson2017automated} to train a BERT model~\cite{devlin-etal-2019-bert}. We then use the trained model to detect hate speech from our data corpus. 

The dataset of Davidson~\etal~\cite{davidson2017automated} contains 24,783 tweets of three categories - \textit{hate speech}, \textit{offensive language}, and \textit{neither}. To facilitate model training and evaluation, the dataset is split, allocating 90\% for the training set and the remaining 10\% for the testing set. We preprocess the dataset by removing stop words using the {\tt wordcloud} package. We then use the {\tt bert\_en\_uncased\_preprocess} model to convert plain text inputs into tokens that are expected by BERT. The classifier is composed of a BERT encoder and an MLP prediction head. In particular, we choose the pre-trained {\tt BERT-Small} model as the encoder, featuring four hidden layers composed of 512 nodes each. We opt for {\tt BERT-Small} because of its capability in achieving \textit{adequate} classification performance, while also being \textit{efficient} in terms of computational requirements. The MLP module consists of three components: a dense layer, a dropout layer (dropout rate $=0.2$)~\cite{srivastava2014dropout}, and another dense layer for predicting labels. We use ReLU activations. The model undergoes a total of 80 epochs. The learning rate is $3\times10^{-5}$. To optimize the training process, we employ the AdamW optimizer~\cite{loshchilov2017decoupled} with a weight decay set to $0$. 

The model achieves an overall accuracy of $91.04$ on the testing set of Davidson~\etal~\cite{davidson2017automated}, suggesting a good performance in hate speech detection. However, it is important to note that although the dataset of Davidson~\etal~\cite{davidson2017automated} provides a valuable resource, the domain of our dataset \textit{may not perfectly align} with theirs. Consequently, any potential domain shift between the two datasets may impact the model's performance when applied to our specific dataset. As a result, we further conduct an experiment to verify the robustness of the trained model on our dataset.

\subsubsection{Robustness Verification}
We sample another validation set of 150 tweets from our dataset. Three researchers read the tweets and independently label them into three categories (\ie hate speech, offensive language, and neither). The final label is assigned with the consensus votes from three annotators. The Fleiss’ Kappa score of the three annotators is 0.35, indicating fair agreement. Subsequently, we evaluate the performance of our classifier using this manually labeled dataset. This three-class classifier achieves an accuracy of 79.33. Moreover, it exhibits a weighted F1 score of 79.59, a precision of 80.96, and a recall rate of 79.33. These results collectively demonstrate a commendable performance for a three-class classification problem. Finally, we apply our model to the entire collected tweets. 

\vspace{-2mm}

\section{Results}
\vspace{-2mm}
From the dataset comprising 503,424 collected tweets, we find 3.33\% ($n=16,784$) of the tweets are classified as hate speech, and 60.11\% ($n=302,605$) are offensive language. The remaining 36.56\% ($n=184,033$) of the tweets are neither hate speech nor offensive language. We remove stopwords and apply lemmatization and tokenization to hate speech tweets. Table~\ref{tab: hate_keywords} shows the top 10 words that appear most frequently in hate speech on NBA athletes. 

\begin{table}[t]
\centering
\caption{Top 10 hate speech keywords related to NBA athletes.}
\label{tab: hate_keywords}
\begin{tabular}{lll}
\hline
Rank & Word   & Frequency \\
\hline
1    & ass    & 1,786      \\
2    & hate   & 1,693      \\
3    & gay    & 801      \\
4    & stupid  & 781      \\
5    & people & 627       \\
6    & white & 617       \\
7    & man & 570       \\
8    & nigger    & 541       \\
9    & dirty  & 463       \\
10   & racist & 436      \\
\hline
\end{tabular}
\end{table}

To identify the NBA athletes who were targeted by hate speech the most, we use the keyword search method. In particular, by leveraging the extracted player names and Twitter handles, we discover the top 50 NBA athletes who are subjected to the highest levels of hateful content. Table~\ref{tab:top_10_hated} shows the top 10 NBA athletes with the most associated hate speech tweets.  Notably, the list of the 50 most hated athletes includes popular names such as Lebron James, Kevin Durant, Ja Morant, Steph Curry, Devin Booker, Anthony Davis, \textit{etc}. Two primary reasons can contribute to the observed phenomenon. Firstly, popular players often attract more attention and discussions, thereby increasing the likelihood of encountering hateful content. The prominence of these players within the NBA creates a higher probability of hate speech directed toward them. Secondly, high-profile players and notable Twitter accounts tend to become targets for hate speech due to the potential for amplified online visibility~\cite{elsherief2018peer}.

 \begin{table}[t]
 \centering
 \caption{Top 10 NBA athletes with the most associated hate speech tweets.}
 \label{tab:top_10_hated}
\begin{tabular}{lll}
\hline
Rank & Player & \# Tweets \\
\hline
1    & Anthony Davis     & 3,211        \\
2    & Ja Morant	     & 2,469        \\
3    &   Anthony Edwards     &   2,173        \\
4    &   Mckinley Wright IV	     &   1,199        \\
5    &   Lonnie Walker IV	     &        1,199   \\
6    &  Alex Len	     &       892    \\
7    &    LeBron James    &      784     \\
8    &   Russell Westbrook	     &  596         \\
9    &   Chris Paul	     &      562     \\
10   &     Kevin Durant	   &       539   \\
\hline
\end{tabular}
\end{table}

To further characterize the targets of hate speech on NBA athletes, we use different sets of keywords to search for relevant tweets. The targeted groups mined are {\tt Black}, {\tt White}, {\tt Jews}, {\tt dirty players}, {\tt LGBTQ}, {\tt chokers},  {\tt selfish players}, {\tt fat players}, {\tt racists}, and {\tt short players}. Table~\ref{tab:keywords} summarizes the keywords used for each group. 

\begin{table}[t]
\centering
\caption{Keywords of the targets of hate speech on NBA athletes.}
\label{tab:keywords}
\begin{tabular}{ll}
\hline
Group          & Keywords                                 \\
\hline
{\tt Black}          & nigger, nigga, n*gg*r, black, niggers    \\
{\tt White}          & white                                    \\
{\tt Jews}           & jews                                     \\
{\tt Dirty player}   & dirty, flop, flopper, flopping           \\
{\tt LGBTQ}          & faggot, gay, lesbian                     \\
{\tt Choker}         & choker, choke artist                     \\
{\tt Selfish player} & selfish, stat padder, ball hog, stat pad \\
{\tt Fat player}     & fat                                      \\
{\tt Racist}         & racist                                   \\
{\tt Short player}   & short, little, small        \\            
\hline
\end{tabular}
\end{table}

The group that experiences the highest degree of targeting is the {\tt Black} community, with a significant count of 4,124 tweets specifically directed toward them. It is worth noting that out of the top 50 NBA athletes that are associated with the most hate speech tweets, 48 are of African descent, while 2 are of Caucasian descent. However, in 2022, approximately 71.8\% of NBA players were African American.\footnote{\url{https://43530132-36e9-4f52-811a-182c7a91933b.filesusr.com/ugd/403016\_901e54ed015c44fb83df939d2070dc17.pdf}} These statistics raise important questions about the potential influence of racial bias in the criticism directed toward athletes.

\begin{figure}[t]
\centering
    \includegraphics[width=0.7\linewidth]{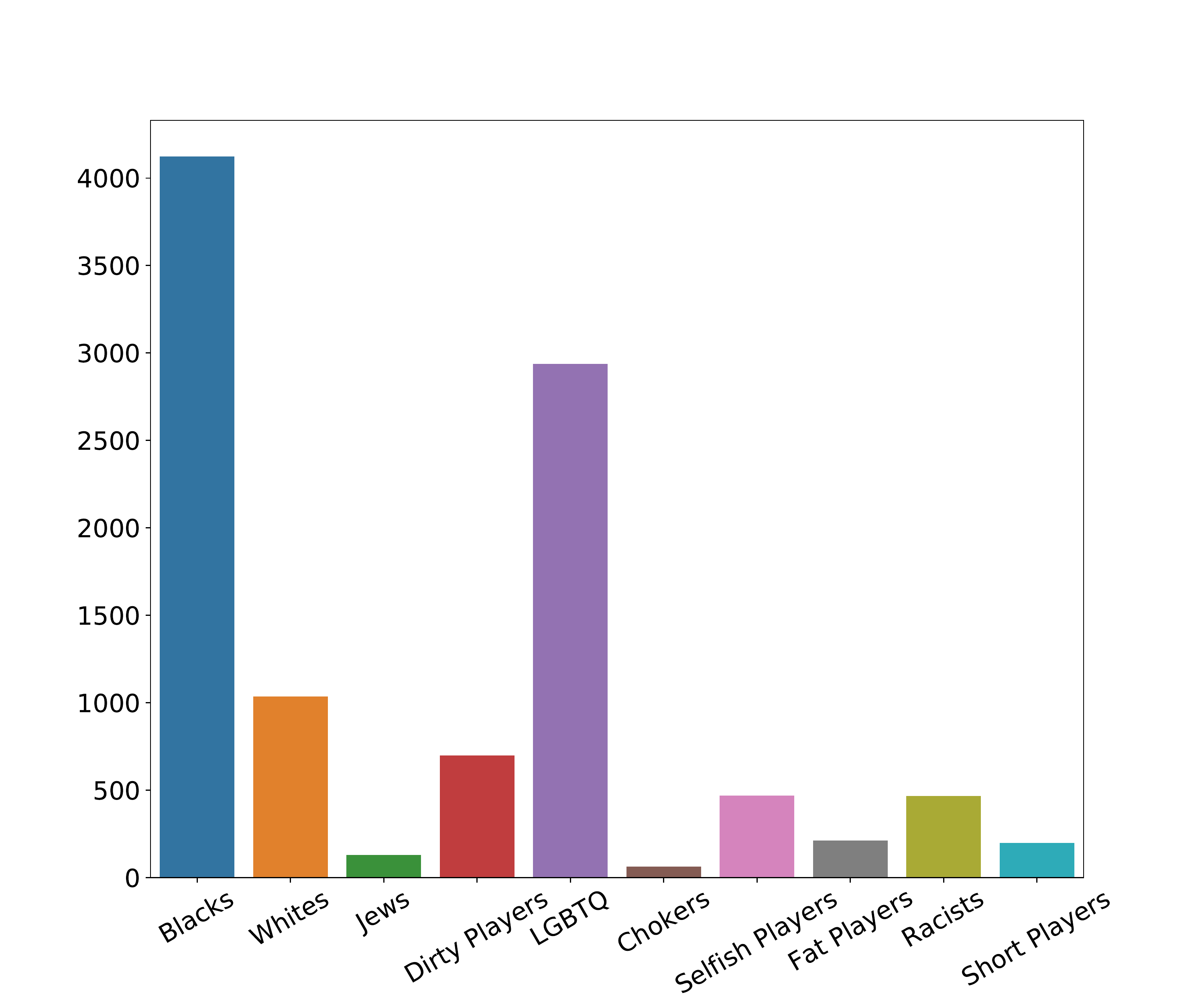}
    \caption{Distribution of tweets related to the targeted groups of hate speech against NBA athletes.}
    \label{fig: Targeted groups distribution}
\end{figure}

Following closely, the {\tt LGBTQ} community faces a substantial number of 2,938 tweets aimed at their community. The count of tweets targeting the {\tt White} individuals ranks third, totaling 1,035 tweets. In fourth place, there are 698 tweets directed toward {\tt dirty players}. Additionally, 470 tweets specifically target {\tt selfish players}, while 468 tweets aim at individuals characterized as {\tt racists}. Moreover, there are 212 tweets targeting {\tt fat players} and 199 tweets focusing on {\tt short players}. The {\tt Jewish} community is the subject of 130 tweets, and 64 tweets are directed at individuals referred to as {\tt chokers}. Figure~\ref{fig: Targeted groups distribution} shows the tweet distribution of targeted groups.

\begin{figure}[h]
\centering
    \includegraphics[width=0.7\linewidth]{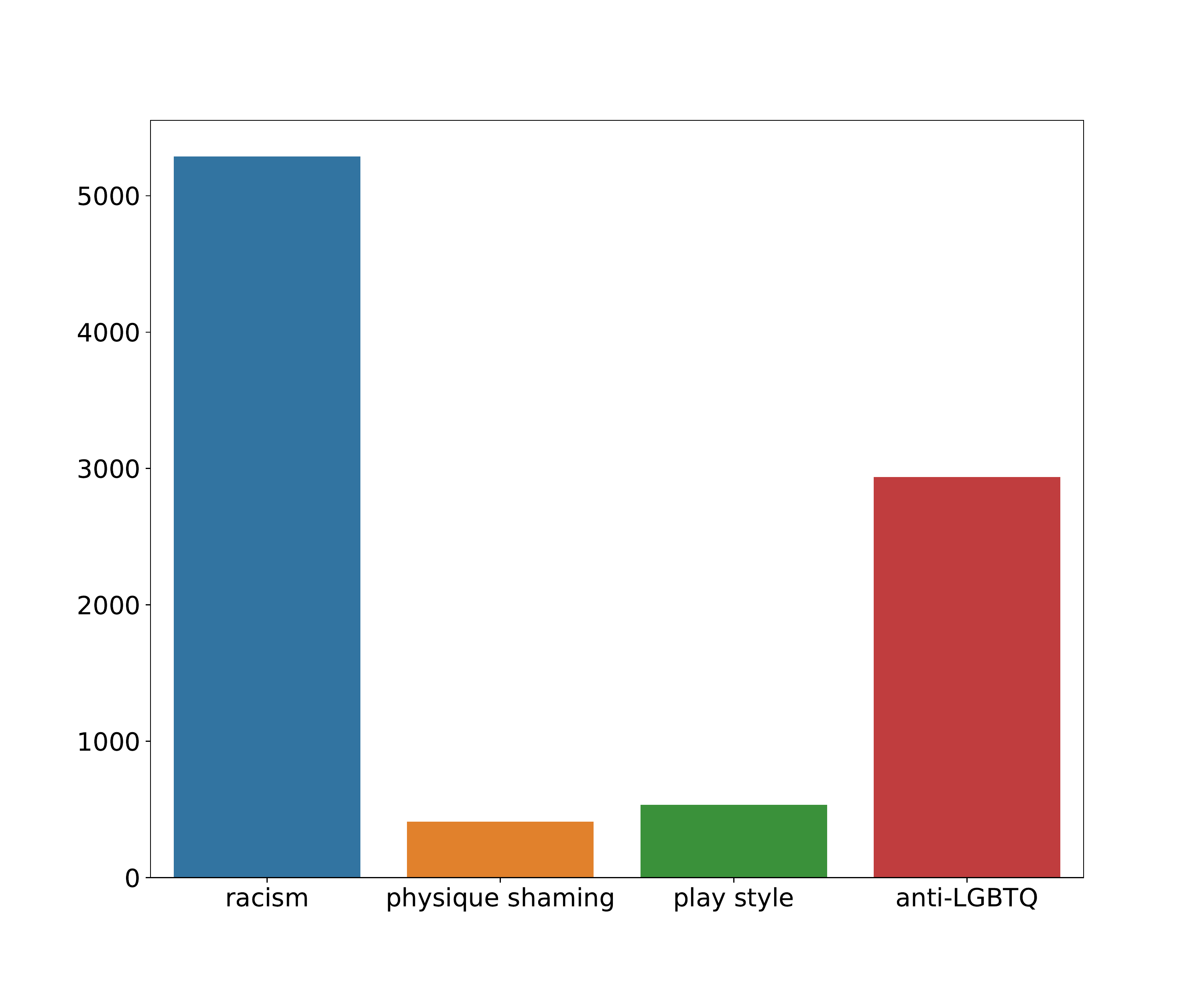}
    \caption{Topic distributions of hate speech tweets related to NBA athletes.}
    \label{fig: Topic distribution}
\end{figure}

Upon identifying the targeted groups within the hate tweets, these categories are subsequently organized into distinct topics, namely racism, physique shaming, play style, and anti-LGBTQ sentiments. More specifically, tweets about {\tt Black}, {\tt White}, and {\tt Jews} are grouped into the racism topic. Tweets about {\tt fat players} and {\tt short players} are included in the physique shaming topic. The play style topic contains tweets about {\tt selfish players} and {\tt chokers}. Tweets about {\tt LGBTQ} are included in the anti-LGBTQ topic. This classification enables a more comprehensive understanding of the underlying themes present within hate speech.  The topic that emerges as the most prevalent is racism, with a support count of 5,289 instances. Following closely, the topic of anti-LGBTQ exhibits a support count of 2,940. Play style, on the other hand, garners a support count of 534, while physique shaming records a support count of 411. These figures highlight the relative prominence and occurrence of each topic within the analyzed hate speech tweets. Figure~\ref{fig: Topic distribution} shows the distribution of these topics.  

To understand the potential correlation between hate speech tweets and players' performance, we compute the correlation coefficients of the number of hate speech tweets and a series of players' performance statistics, player demographics as well as their physical characteristics. The performance statistics of the NBA athletes are collected from \url{basketballreference.com}. Variables include:
\begin{itemize}
    \item \textbf{Age}
    \item\textbf{G: Games Played.} The number of games in which a player has participated.
    \item \textbf{GS: Games Started.} The number of games in which a player was listed as a starter in the team's lineup.
    \item \textbf{MP: Minutes Played.} The number of minutes a player has been on the court during games.
    \item \textbf{TOV: Turnovers.} The number of times a player loses possession of the ball to the opposing team through errors such as bad passes, mishandling the ball, or offensive fouls.
    \item \textbf{Impact:} A player's influence or effect on the game. It encompasses various aspects of a player's performance that contribute to their team's success. The impact of a player can be evaluated through a combination of statistics, observations, and contextual analysis.
    \item \textbf{TS\%: True Shooting Percentage.} It measures a player's shooting efficiency by taking into account their field goals, three-pointers, and free throws.
    \item \textbf{Usage:} It is a metric that quantifies the percentage of team plays or possessions that a player uses while they are on the court. Usage rate helps evaluate the level of involvement and offensive responsibility a player has within their team's offensive system.
    \item \textbf{BMI: Body Mass Index.} It is a measure used to assess body composition and provide an indication of whether a person's weight is within a healthy range relative to their height.
\end{itemize}

\begin{figure}[h]
\centering
    \includegraphics[width=0.7\linewidth]{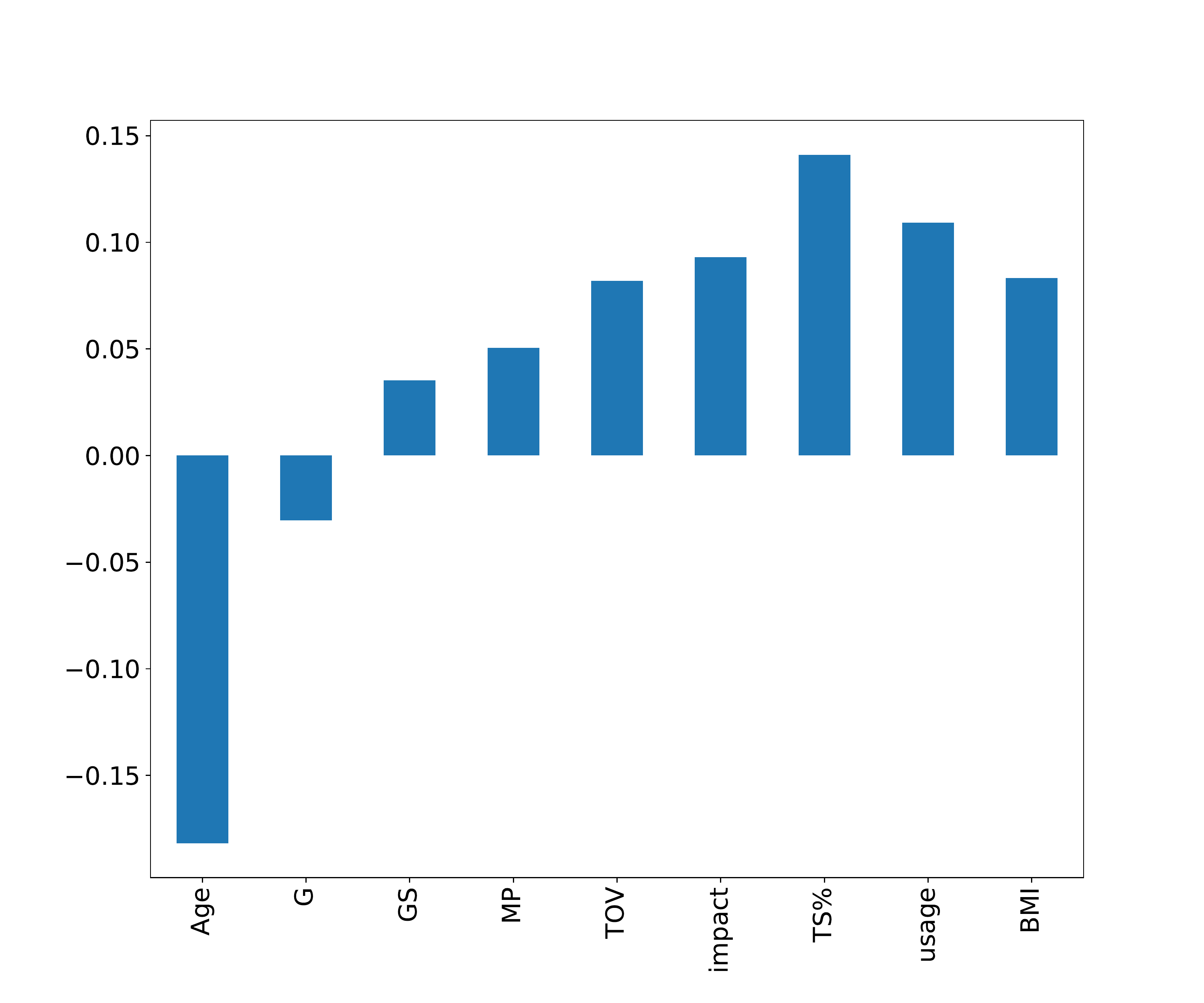}
    \caption{Correlation coefficients between the number of hate speech tweets and variables including performance statistics, demographics, and physical characteristics of the top 50 most hated NBA athletes.}
    \label{fig: correlationbar}
\end{figure}

The results revealed that the number of hate tweets demonstrated positive correlations with GS (Games Started), MP (Minutes Played), TOV (Turnovers), Impact, TS\% (True Shooting Percentage), usage, and BMI (Body Mass Index). Conversely, hate tweet frequency showed negative correlations with age and G (Games Played) (Figure~\ref{fig: correlationbar}).

\begin{figure}[h]
\centering
    \includegraphics[width=0.7\linewidth]{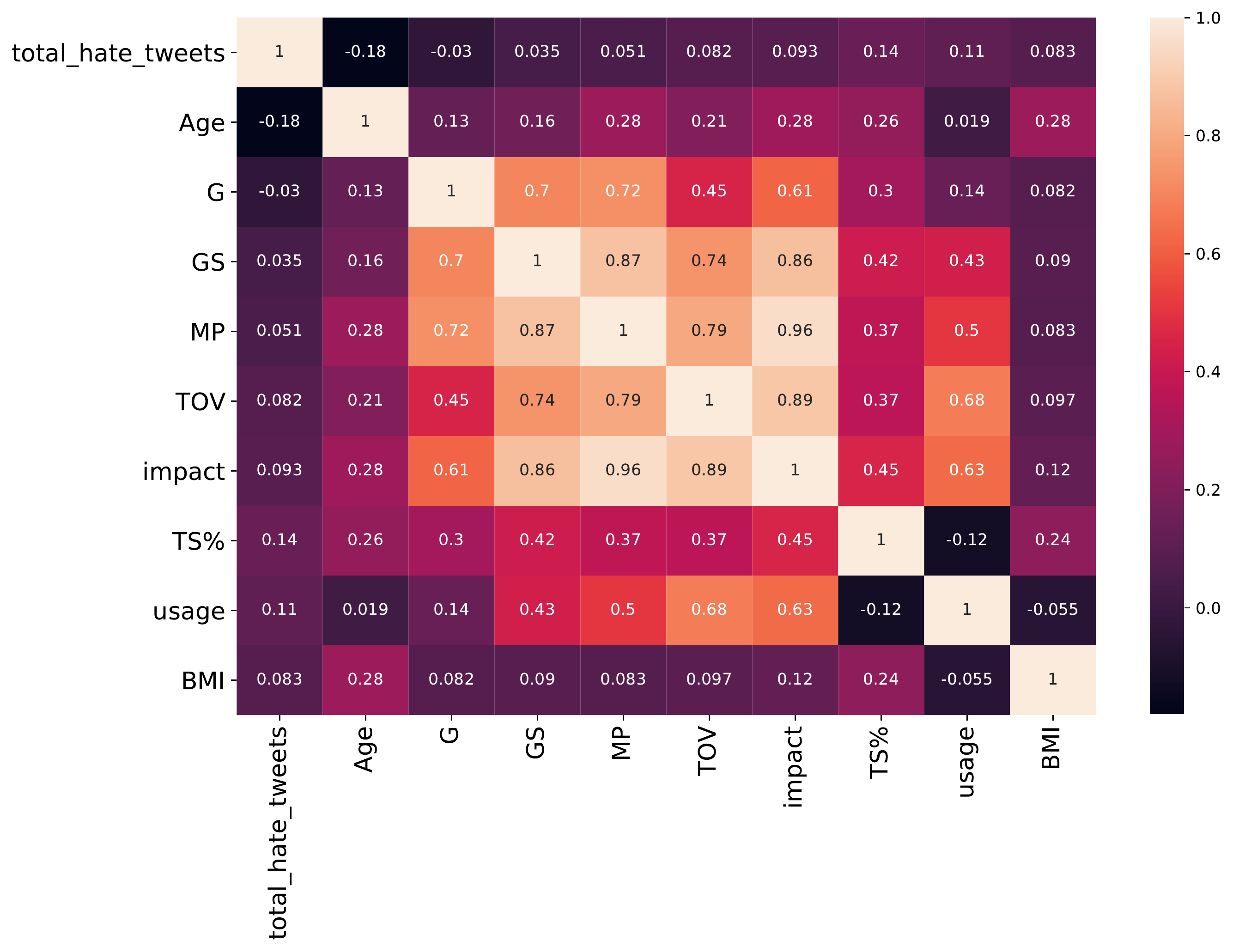}
    \caption{Heatmap of correlations between the attributes of top 50 hated NBA athletes.}
    \label{fig: heatmap}
\end{figure}

However, it is worth noting that MP (Minutes Played), TOV (Turnovers), Impact, GS (Games Started), TS\% (True Shooting Percentage), and usage exhibit strong correlations with each other (Figure~\ref{fig: heatmap}). This suggests that their correlations with the number of hate tweets may be attributed to the fact that they are all performance metrics. Our analysis reveals that players who excel in their performance often become targets of hate speech, likely stemming from rival fans and individuals who may have financial stakes in outcomes, such as bettors.

Regarding the positive correlation observed between BMI and the number of hate speech tweets, we discover that a significant portion of the top 50 most hated NBA athletes consists of individuals categorized as overweight. Specifically, among these athletes, 17 individuals have a BMI exceeding 25. This suggests that their weight status might make them susceptible targets for fat shaming or height shaming through hate speech on social media platforms.

\vspace{-3mm}

\section{Discussions and Conclusions}
\vspace{-3mm}
In this study, we compile a list of hate speech-related and NBA athletes-related keywords to collect tweets that potentially contain hateful content toward NBA athletes. We then fine-tune a BERT model to classify collected tweets into hate speech, offensive language, and neither on an open hate speech dataset~\cite{davidson2017automated}. After examining the classifier performance on a manually labeled subset of our collected tweets, we find that out of the 503,424 tweets, 3.33\% ($n=16,784$) are classified as hate speech, and 60.11\% ($n=302,605$) are offensive language. Our model achieves an overall accuracy of 79.33, and a weighted F1 score of 79.59. These results demonstrate the effectiveness of our classification approach in discerning hate speech and offensive language within the collected dataset.

To gain a deeper understanding of the specific groups that are more susceptible to hate speech, we use the keyword search method. Through this process, we uncover notable patterns indicating that athletes belonging to the {\tt Black} community and the {\tt LGBTQ} community are disproportionately targeted with hate speech. Additionally, players who possess a distinct play style, as well as those who are shorter or overweight, emerge as prominent targets for such abuse. These findings shed light on the specific demographics and characteristics of athletes who are most likely to face hate speech within the NBA community. Racism, physical shaming, play styles, and anti-LGBTQ remarks are the major themes found in our collected dataset.

In conclusion, this study provides valuable insights into the prevalence of hate speech directed toward NBA athletes on social media platforms. By employing a combination of keyword searches and machine learning techniques, we have identified the targeted groups and major themes of hate speech within the NBA community.

Moving forward, further research can explore the impact of hate speech on the mental well-being of the targeted athletes and evaluate potential interventions to mitigate this issue. Additionally, analyzing the role of social media platforms and their policies in addressing hate speech toward athletes could contribute to fostering a safer online environment. It is essential to continue monitoring and addressing this ongoing problem to promote respect, inclusivity, and support for athletes across all platforms. By understanding hate speech in sports and taking proactive measures, we can work toward creating a positive and supportive environment for athletes to thrive both on and off the court.

\bibliographystyle{splncs04}
\bibliography{mybibliography}

\end{document}